\documentclass[12pt]{article}
\pdfoutput=1
\usepackage{a4wide}
\usepackage{amssymb}
\usepackage{amsmath}
\usepackage{graphicx}
\usepackage{mathdots}
\usepackage{slashed}
\usepackage{color}

\def\makeatletter{\catcode`\@=11}% 11:letter
\makeatletter
\def\mathbox#1{\hbox{$\m@th#1$}}%
\def\math@ccstyles#1#2#3#4#5#6#7{{\leavevmode
      \setbox0\mathbox{#6#7}%
      \setbox2\mathbox{#4#5}%
      \dimen@ #3%
      \baselineskip\z@\lineskiplimit#1\lineskip\z@
      \vbox{\ialign{##\crcr
             \hfil \kern #2\box2 \hfil\crcr
             \noalign{\kern\dimen@}%
             \hfil\box0\hfil\crcr}}}}
\def\mathaccstyles{\math@ccstyles\maxdimen}
\def\maththroughstyles{\math@ccstyles{-\maxdimen}}
\def\unity%
 {\maththroughstyles{.45\ht0}\z@\displaystyle {\mathchar"006C}\displaystyle 1}
\makeatletter
\@addtoreset{equation}{section}
\makeatother

\begin{document}

\begin{titlepage}
\begin{center}

\hfill FPAUO-13/12  \\

\phantom{xx}
\vskip 0.4in

{\large \bf Hints of 5d Fixed Point Theories from Non-Abelian T-duality}

\vskip 0.4in

{\bf Yolanda Lozano}${}^{a}$,~{\bf Eoin  \'O Colg\'ain}${}^{b}$, {\bf Diego Rodr\'{\i}guez-G\'omez}${}^{c}$
\\

\vskip .2in

Department of Physics,  University of Oviedo,\\
Avda.~Calvo Sotelo 18, 33007 Oviedo, Spain\\

\vskip .2in

\vskip 0.5in

\end{center}

\centerline{\bf Abstract}

\vskip .2in

\noindent
In this paper we investigate the properties of the putative 5d fixed point theory that should be dual, through the holographic correspondence, to the new supersymmetric $AdS_6$ solution constructed in \cite{Lozano:2012au}. This solution is the result of a non-Abelian T-duality transformation on the known supersymmetric $AdS_6$ solution of massive Type IIA. The analysis of the charge quantization conditions seems to put constraints on the global properties of the background, which, combined with the information extracted from considering probe branes, suggests a 2-node quiver candidate for the dual CFT.

\vfill
\noindent
 {%\footnotesize
 $^a$ylozano@uniovi.es,
$^b$ocolgain@gmail.com,
$^c$d.rodriguez.gomez@uniovi.es}

\end{titlepage}
\vfill
\eject

\tableofcontents

\section{Introduction}

Gauge theories in five dimensions are naively non-renormalizable and hence do not define complete quantum theories \textit{per se}. However, under some circumstances, they can be at fixed points, which can in turn exhibit rather exotic properties such as enhanced global symmetries of exceptional type  \cite{Seiberg:1996bd}.

Minimal supersymmetry in 5d contains 8 supercharges rotated by an $SU(2)_R$ R - symmetry and exactly like in 4d the whole theory on the Coulomb branch follows from a prepotential. However the existence of Chern-Simons terms in five dimensions, together with the 5d analogue of parity anomaly, which generates a CS term upon integrating out massive fermions, crucially constrains the form of the prepotential, allowing in fact to write it exactly. It then turns out that upon suitably choosing the gauge group and matter content one can remove the dimensionful bare coupling and obtain a fixed point theory. Note that even though the theories do not have tunable parameters, they exist for a range of ranks such that a large $N$ limit can be defined \cite{Intriligator:1997pq}. 

A crucial feature of gauge theories in five dimensions is that, for each vector multiplet, we immediately have a topologically conserved symmetry under which instanton particles are electrically charged. These instanton particles have a mass proportional to the Coulomb branch modulus and hence become massless on the Higgs branch. String theory arguments allow, in certain cases, to identify such Higgs branch with the moduli space of  $E$-type instantons, with instanton particles playing a crucial role in such an identification.

Given the existence of fixed point theories admitting a large $N$ limit in five dimensions, it is natural to wonder whether a gravity dual exists in the context of the AdS/CFT correspondence. Indeed, the answer turns out to be positive. Type I' string descriptions dual to five dimensional supersymmetric fixed points with $E_{N_f+1}$ global symmetry were constructed in 
\cite{Brandhuber:1999np}. More general 5d gauge  theories, in particular of quiver type, have also been constructed using the correspondence \cite{Bergman:2012kr,Bergman:2012qh}. Although field-theoretic considerations seem to suggest that quiver gauge theories cannot be at fixed points, as the Coulomb branch moduli space would develop singularities at finite distance, one could expect that instantons becoming massless 
at these singularities could in fact resolve them. While a complete field theory picture for this resolution is yet not available --see nevertheless \cite{Bergman:2013aca}--, it is easy to construct well-behaved $AdS_6$ geometries which should be dual to these theories. Hence, using the gravity dual as a guiding principle we can conclude that quiver gauge theories can also be at fixed points \cite{Bergman:2012kr,Bergman:2012qh}.

In fact, along these lines, the AdS/CFT correspondence can be used as a tool to search for potentially new classes of 5d fixed point theories. Although, perhaps not surprisingly, as one would expect the landscape of 5d CFTs to be more rigid than its 4d analogue, this route reaches a dead end if one looks for ``standard" $AdS_6\times M$ warped solutions in Type IIA \cite{Passias:2012vp}\footnote{An obvious point not discussed in \cite{Passias:2012vp} is that Hopf T-duality leads to a supersymmetric solution in Type IIB.},
a new $AdS_6$ Type IIB background can be constructed \cite{Lozano:2012au} if one
allows for more exotic types of solutions. This new $AdS_6$ solution arises as the result of a 
non-Abelian T-duality transformation of the known supersymmetric $AdS_6$ solution of massive Type IIA  \cite{Brandhuber:1999np}, and should be relevant for defining new classes of 5d fixed point theories through the holographic correspondence. Alternatively, it can be regarded as the supersymmetric vacuum of a 
full embedding \cite{Jeong:2013jfc} of Romans' F(4) gauged supergravity \cite{Romans:1985tw} in Type IIB supergravity. 

As opposed to its Abelian counterpart, non-Abelian T-duality is much less understood. Given the well-documented difficulties in extending Abelian transformations based on Kramers-Wannier duality \cite{Kramers:1941kn} to non-Abelian settings (see for example \cite{Savit:1979ny}) it is rather surprising that there is an immediate generalisation \cite{de la Ossa:1992vc} of the Buscher procedure \cite{Buscher:1987sk, Buscher:1987qj}, or, more precisely, of the gauging procedure derived by Rocek and Verlinde thereof \cite{Rocek:1991ps}. Still, long-standing questions remain, notably the status of the transformation as a string theory symmetry and the fate of global aspects  \cite{Giveon:1993ai, Alvarez:1993qi, Alvarez:1994np}. Indeed, while Abelian T-duality maps an $S^1$ to another $S^1$ with inverse radius, it is not known what ranges one should attribute to the dual coordinates of an $S^3$ under non-Abelian T-duality. Despite these open problems, recently we have witnessed a small resurgence in interest in non-Abelian T-duality, spurred on by the extension of non-Abelian T-duality to incorporate RR fields \cite{Sfetsos:2010uq, Lozano:2011kb}.\footnote{See also \cite{Gevorgyan:2013xka} for the role the Fourier-Mukai transform plays in the transformation of the RR fields.}
Provided one is careful about the R symmetry, it is possible to generate supersymmetry preserving  solutions  of relevance in the context of gauge/gravity duality
\cite{Lozano:2012au,Itsios:2012zv,Itsios:2013wd} as well as to study the implications for G-structures \cite{Barranco:2013fza,Macpherson:2013zba}. Reversing the logic we will see that it is also possible to extract some global information by analyzing the implications of the AdS/CFT correspondence on the newly generated background.

The paper is organized as follows. 
We start in section 2 by summarizing the main properties of the supersymmetric $AdS_6$ solution of massive Type IIA constructed in  \cite{Brandhuber:1999np}. In section 3 we present the non-Abelian T-dual background, expanding on the results in \cite{Lozano:2012au}.  In section 4 we discuss the different charges present in the dual background, which in order to be properly quantized requires a specific global completion of the dual geometry. This completion leads to an interpretation in terms of a D5-D7 system, but raises some concerns on the dual geometry that we discuss. The possibility of a non-compact dual space is also analyzed. This implies however the existence of a continuous spectrum of fluctuations through the spherical Bessel function.
In section 5 we analyze some of the properties of the 5d CFT that should be dual to the new $AdS_6$ background through the holographic correspondence. We explore the Coulomb branch of the theory as well as its  instanton and baryon vertex configurations. These allow for a concrete proposal for a 5d CFT in terms of two gauge groups and two flavor symmetries. We comment on the apparent non-existence of a Higgs branch.  In section 6 we calculate the entanglement entropy of the dual background. This implies an $S^5$ free energy for the 5d dual CFT that differs from that of the original theory. Section 7 contains some Discussion of the open problems left out by our analysis. Appendix A contains a detailed analysis of the supersymmetry properties of the Hopf T-dual of the original background. Appendix B contains the supersymmetry analysis of the non-Abelian T-dual background. Appendix C complements the construction of BPS probe branes made in section 5 with a kappa symmetry analysis.

\section{The D4-D8 brane system}
 
The Coulomb branch of supersymmetric 5d gauge theories is completely contained in a prepotential severely constrained by the existence of Chern-Simons terms. Inspection of the prepotential shows that for a $USp(2\,N)$ gauge theory with one antisymmetric hypermultiplet and $N_f<8$ fundamental hypermultiplets the bare coupling can be safely removed by taking it to infinity. The theory is therefore expected to be a strongly coupled fixed point theory. On the other hand, this theory can be engineered in string theory on a stack of $N$ D4 branes probing a $O8^-$ plane with $N_f$ coincident D8 branes. Conversely, one can find a massive IIA solution corresponding to this Type I' configuration which, in the near-brane region, becomes the warped product of $AdS_6$ times a half-$S^4$. The corresponding background is (we use the conventions in \cite{Lozano:2012au})

\begin{eqnarray}
\label{IIAbackground}
ds^2& = & \frac{W^2\,L^2}{4}\,\Big[9\,ds^2(AdS_6)+4\,ds^2(S^4)\Big] \nonumber \\
F_4&=&5\,L^4\,W^{-2}\,\sin^3\theta\,d\theta\wedge {\rm Vol}(S^3)\nonumber \\
e^{-\phi}&=&\frac{3\,L}{2\,W^{5}}\, ,\qquad W=(m\,\cos\theta)^{-\frac{1}{6}}
\end{eqnarray}
where $m$ is the Romans' mass, $m=(8-N_f)/(2\pi)$ (we take $l_s=1$),
$L$ denotes the $AdS_6$ radius  
and the metric on $S^4$ takes the form 
\begin{eqnarray}
\label{theS^3}
ds^2(S^4) &=&  d \theta^2 + \sin^2 \theta\, ds^2(S^3). 
\end{eqnarray}
While $S^4$ would have $SO(5)$ isometry, the $\theta$-dependent warping means that this is broken to $SO(4) \sim SU(2) \times SU(2)$. Upon dimensional reduction, the $SU(2)\times SU(2)$ isometry leads to two gauge fields in $AdS_6$, standing for the global symmetries of the dual CFT. One of the $SU(2)$ corresponds to the $SU(2)_R$ R-symmetry of the field theory and the other one to the $SU(2)_M$ mesonic symmetry acting on the antisymmetric hypermultiplet. Besides, there is an extra Abelian gauge field in $AdS_6$ coming from the RR 1-form potential, which stands for the global instantonic symmetry of the dual CFT.

Writing the geometry in terms of $\alpha=\frac{\pi}{2}-\theta$, the $O8^-$ action involves an inversion of the transverse coordinate, which translates, in the near-brane region, to $\alpha\rightarrow -\alpha$. Hence the range of $\alpha$, which would naively be $[-\frac{\pi}{2},\,\frac{\pi}{2}]$ in order to cover the full $S^4$, is reduced to $[0,\,\frac{\pi}{2}]$ upon modding. This corresponds to $\theta\in[0,\, \frac{\pi}{2}]$. The $O8^-$ location at $\alpha=0$ becomes $\theta= \frac{\pi}{2}$, where the dilaton diverges.
This is the reflection on the gravity side  of  the removal of the dimensionful bare gauge coupling which puts the field theory at a fixed point. Indeed, resorting to the full string theory picture to resolve the singularity, upon tuning the dilaton to diverge right on top of the orientifold, the $N_f$ D8 branes on top of the $O8^-$ give rise to the enhanced $E_{N_f+1}$ global symmetry \cite{Seiberg:1996bd}.

For a generic value of $\theta$ away from the orientifold singularity, curvature and dilaton go like
\begin{equation}
\mathcal{R}\sim-\frac{m^{\frac{1}{3}}}{L^2}\,, \qquad  e^{\phi}\sim \frac{1}{L\,m^{5/6}}\, .
\end{equation}
Hence, in order to ensure the validity of the solution we need to demand
\begin{equation}
\frac{m^{\frac{1}{3}}}{L^2}<<1\, , \qquad L\,m^{5/6}>>1\, .
\end{equation}
Being $m=(8-N_f)/(2\pi)$ the Romans mass quantization condition these conditions simply reduce to $L >> 1$. Using the correct quantization of the four-form flux, one gets
\begin{equation}
\label{4formfluxN} 
L^4\,m^{1/3} = \frac{16 \pi}{9}  N, 
\end{equation}  
where $N>>1$ stands for the number of D4 branes. 

\subsection{Supersymmetry}

Writing the metric on $S^3$ in terms of a Hopf-fibre over $S^2$, 
\begin{equation}
ds^2(S^3) = \frac{1}{4} \left[ d \phi_1^2 + \sin^2 \phi_1 d \phi_2^2 + (d \phi_3 + \cos \phi_1 d \phi_2)^2 \right],
\end{equation}
omitting details, the Killing spinors take the form 
\begin{equation}
\label{mIIA_KS}
\eta = (\cos \theta)^{-1/12}  e^{-\frac{\theta}{2} \gamma \Gamma^{\theta} \sigma^1} e^{-\frac{\phi_1}{2} \Gamma^{\phi_3 \phi_2} } e^{-\frac{\phi_2}{2} \Gamma^{\phi_2 \phi_1} } \tilde{\eta}, 
\end{equation}
where  $\gamma = \Gamma^{\theta \phi_1 \phi_2 \phi_3}$ and $\tilde{\eta}$ denotes the Killing spinor on $AdS_6$. $\eta$ is subject to a single projection condition 
\begin{equation}
\label{ads6s4proj1}
\left[ \sin \theta \Gamma^{\theta} \sigma^1 + \cos \theta \Gamma^{\theta \phi_1 \phi_2 \phi_3}  \right] \eta = - \eta, 
\end{equation}
so there are sixteen supersymmetries, the minimum required for a supersymmetric $AdS_6$ geometry. Furthermore, as is evident from the explicit form of the Killing spinor, it is independent of $\phi_3$, so that if one performs an Abelian T-duality transformation along this direction no supersymmetries will be broken. We show this in detail in Appendix A.

\section{The $AdS_6$ non-Abelian T-dual}

Non-Abelian T-duality with respect to any of the $SU(2)$ subgroups of the $SO(4)$ isometry group of the $S^3$ contained in the internal space of the previous $AdS_6$ background
produces yet another $AdS_6$ solution, this time in Type IIB, which exhibits an explicit $SU(2)$ symmetry   \cite{Lozano:2012au}. The reduction of global symmetries under non-Abelian T-duality is a generic feature, with the isometries being dualized typically being destroyed in the duality. Using spherical coordinates adapted to the remaining $SU(2)$ symmetry the space dual to the $S^3$ is locally $\mathbb{R}\times S^2$. Its global properties are however mostly unknown, this being related to our lack of knowledge on how to extend the gauging procedure used to construct the non-Abelian T-dual \cite{de la Ossa:1992vc} to topologically non-trivial worldsheets \cite{Alvarez:1993qi}.

The $AdS_6$ non-Abelian T-dual constructed in \cite{Lozano:2012au} is given by

\begin{eqnarray}
\label{NATD}
ds^2&=&\frac{W^2\,L^2}{4}\,\Big[\,9\,ds^2(AdS_6)+4\,d\theta^2\,\Big]+e^{-2\,A}\,dr^2+\frac{r^2\,e^{2\,A}}{r^2+e^{4\,A}}\,ds^2(S^2) \nonumber \\
B_2&=&\frac{r^3}{r^2+e^{4\,A}}\, {\rm Vol}(S^2)\, \qquad e^{-\phi}=\frac{3\,L}{2\,W^5}\,e^A\,\sqrt{r^2+e^{4\,A}} \nonumber \\
F_1&=&-G_1-m\,r\,dr\qquad F_3=\frac{r^2}{r^2+e^{4\,A}}\,[-r\, G_1+m\, e^{4A} \,dr]\wedge {\rm Vol} (S^2)
\end{eqnarray}
\noindent with $ds^2 (S^2)$,  ${\rm Vol}(S^2)$ given by

\begin{equation}
ds^2 (S^2)=d\xi^2+\sin^2\xi\,d\psi^2\, , \qquad {\rm Vol}(S^2)=\sin\xi\,d\xi\wedge d\psi\, , 
\end{equation}
and
\begin{equation}
\label{G1}
e^A=\frac{W\,L}{2}\,\sin\theta\, , \qquad G_1=\frac{5}{8\,W^2}\,L^4\,\sin^3\theta\,d\theta\, .
\end{equation}

For later purposes, the Hodge-dual RR field strengths are given by

\begin{equation}
F_9=\frac{3^6}{2^6}\, W^3\, L^7 \frac{r^2\, e^A\,} {r^2+e^{4A}}\left[ \frac58\, L^2\, \sin^3{\theta}\, dr-
m\, r\, e^{2A}\, W^4\, d\theta\right] \, \wedge d{\rm Vol}(AdS_6)\wedge {\rm Vol}(S^2)
\end{equation}
\noindent and
\begin{equation}
F_7=-\frac{2^6}{3^6} W^3 L^7 \left[ \frac58 \,r\, e^{-3A}\, L^2\, \sin^3{\theta}\, dr+m\, e^{3A}\, W^4\, d\theta\right]  \wedge d{\rm Vol}(AdS_6)\, .
\end{equation}

In \cite{Itsios:2012dc} the non-Abelian T-dual of a general class of Type II supergravity solutions with isometry $SO(4)\sim SU(2)\times SU(2)$ was generated and shown to satisfy the supergravity equations of motion. These results guarantee that (\ref{NATD})  satisfies the Type IIB equations of motion for any positive value of $r$. In order to fully clarify the nature of the space spanned by dual variables one needs to resort to the sigma-model derivation of the transformation. As we have said, no global properties can however be inferred from it in the non-Abelian case. 
We will assume in what follows that $r\in [0,\,R]$ for some regulator $R$ which might be taken to infinity, and try to infer global properties by demanding consistency to the dual background.

In addition to the singularity at $\theta=\frac{\pi}{2}$, inherited from the original background, there is a second singularity at $\theta=0$. This happens because the $S^3$ shrinks to zero size and is completely analogous to the singularity that appears after Abelian T-duality on a shrinking circle.
One can also check that the curvature invariants for this geometry are perfectly smooth for all $r\in [0,\,\infty)$. 

Close to $r=0$ the metric looks like
\begin{equation}
ds^2=h^{-\frac{1}{2}}\,\Big[\,9\,ds^2(AdS_6)+4\,d\theta^2\,\Big]+h^{\frac{1}{2}}\,\,\Big[\frac{dr^2+r^2\, ds^2(S^2)}{\sin^2{\theta}}\Big]\, ;\qquad h^{-\frac{1}{2}}=\frac{W^2\,L^2}{4}\, ,
\end{equation}

\noindent so locally the transverse space becomes just $\mathbb{R}^3$. The curvature is in turn given by
\begin{equation}
\mathcal{R}=-\frac{m^{\frac{1}{3}}}{L^2}\,\frac{(29+25\,\cos2\theta)}{3\,\cos^{5/3}\theta\,\sin^2\theta}
\end{equation}
and the dilaton 
\begin{equation}
e^{\phi}=\frac{16}{3\,L^4\,(m\cos{\theta})^{1/3}\sin^3\theta}\sim \frac{1}{L^4\,m^{1/3}}\, .
\end{equation}
Therefore the solution is valid when
\begin{equation}
\frac{m^{\frac{1}{3}}}{L^2}<<1\, , \qquad L^4\,m^{1/3}>>1\, .
\label{cond}
\end{equation}
At large $r$ we find a geometry of the form
\begin{equation}
\label{larger}
ds^2=h^{-1/2}\,\Big[\,9\,ds^2(AdS_6)+4\,d\theta^2+\sin^2\theta\,ds^2(S^2) \,\Big]+h^{1/2}\,\Big[\frac{dr^2}{\sin^2\theta}\Big]\, ; \qquad h^{-\frac{1}{2}}=\frac{W^2\,L^2}{4}
\end{equation}
with a curvature
\begin{equation}
\mathcal{R}=-\frac{m^{\frac{1}{3}}}{L^2}\,\frac{(69+4\,\cos2\theta-\cos4\theta)}{18\,\cos^{5/3}\theta\,\sin^2\theta}
\end{equation}
and a dilaton

\begin{equation}
e^{\phi}=\frac{4}{3\,L^2\,r\,\sin\theta \, (m\cos{\theta})^{2/3}}\,e^{4A}\sim\frac{1}{L^2\,m^{2/3}\,r}\, .
\end{equation}
The same conditions (\ref{cond}) ensure that both dilaton and curvature remain under control, away of course from the known singularities at $\theta=0,\,\frac{\pi}{2}$. The geometry spanned by $\theta$ and $S^2$ is conformally a singular cone at $\theta=0$ with an $S^2$ boundary.

%Note that, given that  $\theta$ terminates at $\pi/2$, the large $r$ geometry looks like a cone with an $S^2$ boundary. }

The fact that the dual geometry is perfectly well-defined for all $r$ leads to a puzzle for finite $R$,
as the geometry would be terminating at a smooth point. We postpone a more detailed discussion of this issue to later sections.

\subsection{Supersymmetry} 

As shown in \cite{Lozano:2012au} this new $AdS_6$ background provides the first example of a non-Abelian T-dual geometry with supersymmetry fully preserved. Appendix B contains the detailed analysis supporting this statement.
To this end, we follow arguments presented in \cite{Itsios:2012dc} and demonstrate that the effect of an $SU(2)$ transformation for space-times with $SO(4)$ isometry is simply a rotation on the Killing spinors. The calculations presented in Appendix B generalise the analysis of \cite{Itsios:2012dc} to include transformations from massive IIA to Type IIB supergravity and provide other necessary details. Once again the key observation will be that there is a rotation of the Killing spinor \cite{Itsios:2012dc} that allows to recast the Killing spinor equations for the T-dual geometry in terms of the Killing spinor equations of the original geometry.

\section{Quantization conditions in the dual theory and the cut-off in $r$}

The RR fluxes of the dual $AdS_6$ background are the gauge invariant fluxes (see  \textit{e.g.} \cite{Grana:2005jc})

\begin{equation}
F_p=d\,C_{p-1}-H_3\wedge C_{p-3}
\end{equation}
satisfying

\begin{equation}
dF_p=H_3\wedge F_{p-2}\, .
\end{equation}
It is well-known  however that the Page charges are the ones that should be quantized \cite{Marolf:2000cb}, although they are non-invariant under large gauge transformations of the $B_2$ field. Large gauge transformations are indeed relevant if 2-cycles exist in the geometry. In view that at least at large $r$ there is a singular cone in the geometry with an $S^2$ boundary, we can explore the implications of large gauge transformations on this $S^2$.
To that matter, recall that the $B_2$ field in (\ref{NATD}) is given by
\begin{equation}
B_2=\frac{r^3}{r^2+e^{4\,A}}\,{\rm Vol}(S^2)\, ,
\end{equation}
so we would expect large gauge transformations 
shifting $B_2$ by $n\,\pi\,{\rm Vol}(S^2)$, with $n$ an integer, 
\begin{equation}
\label{B2}
B_2=\Big(\frac{r^3}{r^2+e^{4\,A}}-n\,\pi\Big)\,{\rm Vol}(S^2)\, ,
\end{equation}
such that around the $S^2$
\begin{equation}
\label{quantB2}
b=\frac{1}{4\,\pi^2}\int\,B_2\,\in\,[0,\,1]\, .
\end{equation}
In the absence of a global definition of the newly generated background it is not clear whether a non-trivial $S^2$ exists at finite $r$. Still, we can consider the $S^2$ in (\ref{larger}) at large $r$.
Then $n$ should be chosen such that
\begin{equation}
\label{B2larger}
B_2\sim \Big(r-n\,\pi\Big)\,{\rm Vol}(S^2)\, ,
\end{equation}
satisfies the quantization condition (\ref{quantB2}). This implies that $n$ should be a function of $r$.
This is somewhat reminiscent of the cascade in \cite{Klebanov:2000hb}, with the important difference that the cascading does not take place in the holographic direction but in an internal direction. As we show below this imposes non-trivial conditions on the dual background.

The Page charges in the dual theory are associated to the currents $d\star\hat{F}=\star\,j^{Page}$ where (see \textit{e.g.} \cite{Benini:2007kg})

\begin{equation}
\hat{F}=F\,e^{-B_2}\, .
\end{equation}
Explicitly in our background we have

\begin{equation}
\hat{F}_1= F_1\, , \qquad  \hat{F}_3 = F_3-B_2\wedge F_1\, ,\qquad  \hat{F}_5=F_5-B_2\wedge F_3+\frac{1}{2}\,B_2\wedge B_2\wedge F_1
\end{equation}
with

\begin{equation}
\label{hatsfields}
\hat{F}_1 =-G_1-m\,r\,dr\, ,  \qquad \hat{F}_3 =\Big((m\,r^2\,-\pi\,m\,n\,r)\,dr-\pi\,n\,\,G_1\Bigr)\wedge {\rm Vol}(S^2)\, , \qquad \hat{F}_5=0\, .
\end{equation}

These fluxes should satisfy  the quantization condition (in $\ell_s=1$ units):

\begin{equation}
\label{quantfluxes}
\frac{1}{2\,\kappa_{10}^2}\,|\int_{\gamma_p} \hat{F}_p | = T_{8-p} \,N_{8-p}\, ,
\end{equation}
along some compact $\gamma_p$ cycle of the dual background\footnote{Note that we added an absolute value so that all our integers will be positive.}.  In the absence of a  clear global definition for the dual background we are going to assume 
 that  non-trivial 1 and 2-cycles exist in the dual geometry and analyse the implications of this on the dual CFT. We will see that indeed many qualitative properties of the original CFT will be reproduced in terms of Type IIB configurations.

Imposing the $\hat{F}_1$ quantization condition we find two integers coming from the two components of $\hat{F}_1$
\begin{equation}
\label{N7}
N_7^r=\frac{m\,R^2}{2}\, , \qquad N_7^{\theta}=\frac{9}{32}\,L^4\,m^{\frac{1}{3}}\, ,
\end{equation}
where the subscripts $r,\,\theta$ refer to the direction that has been integrated. 

The $\hat{F}_3$ quantization conditions lead  to
\begin{equation}
\label{N5}
N_5^r=\frac{m}{\pi}\,\int_0^R\,dr\,(r^2-n\,\pi\,r)\, ,\qquad N_5^{\theta}=\frac{20}{9}\,N_7^{\theta}\,\int_0^{\frac{\pi}{2}} d\theta\,n\,\cos^{\frac{1}{3}}\theta\,\sin^3\theta\, .
\end{equation}
In these integrals $n$ has to be chosen such that the $B_2$ field satisfies (\ref{quantB2}) for each $r$ and $\theta$.
Note that the relative sign in $N_5^r$ seems to imply that this charge will become zero at some point and from then on negative, possibly giving rise to tensionless branes. Nevertheless, taking the implicit $r$ and $\theta$ dependence of $n$ into account, one can see that
$N_5^r$ remains in fact positive for all values of $R$ and $\theta$. 

In order to keep the correct periodicity as one moves in the internal geometry, the $B_2$ field should undergo a sequence of large gauge transformations very reminiscent of the cascade.
The large gauge transformations induce then a change in the Page charges, such that implicitly $N_5^r$ is a function of $\theta$ while $N_5^\theta$ is a function of $r$. Since the charges cannot be integers at the same time for all values of $r$ and $\theta$, the background turns out to be globally inconsistent.  The only way out of this inconsistency is to fix $R$ such that $B_2$ is not allowed to undergo any large gauge transformation, thus forcing $n$ to be zero in (\ref{N5}). Given (\ref{B2larger}), $b$ will be in $[0,\,1]$ as long as 
$r\leq \pi$. It is then natural to fix $R=\pi$, and the Page charges of our background to\footnote{Strictly speaking $R\leq \pi$, but for the sake of concreteness we choose the most natural value $R=\pi$.}
\begin{equation}
\label{allcharges}
N_7^r=\frac{m\,\pi^2}{2}\, ,\qquad N_5^r=\frac{m\,\pi^2}{3}\, , \qquad N_7^{\theta}=\frac{9}{32}\,L^4\,m^{\frac{1}{3}}\, ,\qquad N_5^{\theta}=0. 
\end{equation}

Note that $N_7^r$ and $N_5^r$ are related to the mass of the original Type IIA background and clearly they are not independent once $R$ has been fixed. For $R=\pi$ in particular both charges satisfy $2 N_7^r=3N_5^r$. $N_7^\theta$ on the other hand cannot be an integer if $L$ and $m$ satisfy the conditions given by (\ref{4formfluxN}). This happens because non-Abelian T-duality changes the volume of the dual manifold. 
This change can be absorbed re-defining $\kappa_{10}$, as one does after an Abelian T-duality along a coordinate with periodicity different from  $2\pi$ (see for instance \cite{Breckenridge:1996tt}). One can check that indeed for  a ${\tilde \kappa}_{10}$ satisfying
${\tilde \kappa}_{10}^2=(2\pi)^8 q$, with $q$ an integer such that $N_4=q N_7^\theta$, all dual charges are integers and are related either to the D8-brane charge ($N_7^r$, $N_5^r$) or the D4-brane charge ($N_7^\theta$) of the original background. Consistently, as we will find in the next section when discussing the holographic aspects of the non-Abelian T-dual, $N_7^r$ and $N_5^r$  will be interpreted as flavor charges and $N_7^\theta$  as color charge. Indeed, we will see that there are BPS stable D-branes responsible for these charges independently on the existence of non-contractible cycles in the geometry.

Finally, it is interesting to note that if one generates a $B$-field in a D-brane background as instructed by Ref. \cite{Maldacena:1999mh}, then the Page charge of the induced flux is typically zero. Here $N_5^{\theta} = 0$ is indicating that this charge is induced. 

\subsection{On compact \textit{vs} non-compact $r$}

We have just seen that if $r$ is compact, under reasonable assumptions, the quantization of the Page charges would imply that its maximum value has to be set to $R= \pi$, so as to have a globally well-defined background where all Page charges are integers. Although for
compact $r$ the spectrum of fluctuations is discrete --which is what one would naively expect for an $AdS$ background dual to a CFT, as we will discuss in the next section--, it is very puzzling that
the geometry has to terminate at a perfectly well-defined value at which no invariant quantity blows up.\footnote{Such termination at a regular point would seem to demand the inclusion of extra localized sources to satisfy the equations of motion there. However, motivated by the analysis of the forthcoming sections --to which we refer--, it is tempting to speculate that the non-Abelian T-dual background is sort of an effective description of a complete geometry where no such extra localized objets are present.} The possibility of having tensionless branes beyond $r=R$, suggested by the minus relative sign in the expression for $N_5^r$ in (\ref{N5}), does not allow either for a natural cut-off, given that, as we have mentioned, $N_5^r$ turns out to be strictly positive for all $R$. We will come back to this issue of the termination of the background at a regular point in the next section.
 
One further possibility one might explore is that $R\rightarrow\infty$, so that quantization of the $r$ component of fluxes must not be imposed. However, this possibility raises other concerns inspired in the $AdS/CFT$ correspondence. We have seen that for asymptotically large $r$ the geometry is given by (\ref{larger}), where in particular the $r$ coordinate lives in $\mathbb{R}^+$. Therefore we should expect fluctuations to behave asymptotically as $e^{i\,k\,r}$ for continuous $k$. Since we expect our background to be holographically dual to a 5d fixed point theory (see next section) such fluctuations would be dual to operators in the CFT with conformal dimension proportional to the continuous parameter $k$, which points at a sick dual CFT. We stress that this argument alone, regardless of the quantization conditions for Page charges and the existence of an $S^2$ where we can quantize large gauge transformations of $B_2$, leads to consider, under the light of the $AdS/CFT$ correspondence, a cut-off space.

In order to sustain this qualitative argument one would require to explicitly compute the spectrum of fluctuations in the non-Abelian T-dual $AdS_6$ background. In particular, one might worry that, due to the non-trivial dilaton, warping and fluxes, it might be that somehow the continuous spectrum is avoided. This analysis appears however as a daunting task. To this end, we have performed  a preliminary check on the technically simpler $AdS_3\times S^3\times T^4$ background, that gives some evidence for a continuous spectrum of fluctuations. Therefore, $r$ non-compact raises as well important concerns in the dual theory.

As usual, one should perform a linearised fluctuation analysis around the non-Abelian T-dual solution, along similar lines to seminal studies of the Kaluza-Klein spectra of Freund-Rubin solutions \cite{Duff:1984sv,DeWolfe:2001nz}. This approach runs into a number of difficulties. Firstly, despite the original $AdS_3 \times S^3 \times T^4$ geometry being of Freund-Rubin type, the non-Abelian T-dual is clearly not. Moreover, the internal space is no longer compact, so one also has to work without the usual crutch of the Hodge decomposition theorem that allows one to expand the gauge potentials. 

In the face of these difficulties, our approach will be to work at the non-linear level, borrow intuition from non-Abelian T-duality and at the end linearise the ``fluctuations"\footnote{Strictly speaking these are not fluctuations as we work at the non-linear level. By ``fluctuation", we mean any deformation from the underlying solution, which in this case is $AdS_3 \times S^3 \times T^4$.} by dropping quadratic terms. For simplicity, we will also focus on a single breathing mode, $A$, which can be decoupled and analysed independently from other fluctuations. For concreteness, we consider the following Ansatz 
\begin{equation} 
\label{ansatz}
ds^2 = e^{2 A} \left[ ds^2 (AdS_3) + ds^2 (S^3) \right] + ds^2(T^4)
\end{equation}
where $A$ is a function of the coordinates on both $AdS_3$ and $S^3$, and the latter are normalised so that $R_{\mu \nu} = - \frac{1}{2} g_{\mu \nu}$ and $R_{mn} = \frac{1}{2} g_{mn}$, respectively\footnote{The choice of normalisation follows from \cite{Sfetsos:2010uq,Itsios:2012dc} and the conventions for the type II equations of motion we take from \cite{Itsios:2012dc}. }.  On its own the addition of the breathing mode is not consistent and one needs to support it through complementary fluctuations in order  that the equations of motion are satisfied. We will work on the assumption that this can be done and leave a more detailed analysis to future work. 

The above Ansatz then fits into the class of spacetimes with $SO(4)$ isometry and one can apply the non-Abelian T-duality transformation rules of Ref. \cite{Itsios:2012dc}. Since we are only affecting one $SU(2)$ factor in the T-duality, it is reasonable to expect that singlet KK modes with respect to $SU(2)$ will survive the process. For scalars, such as $A$ above, this means that $A$ should be independent of the coordinates on the $S^3$. This can be easily seen by taking the vector fields dual to either left or right-invariant one forms on $S^3$ and calculating the Lie derivative of $A$ with respect to the vectors. 

Thus, if we want to consider KK modes on $S^3$, there is no way from the offset that we can simply perform non-Abelian T-duality on these modes as they are not singlets. However, as we shall see for the above scalar, it is possible to reconstruct the spectrum, at least when one simply focusses on the dilaton equation 
\begin{equation}
\label{dil}
R + 4 \nabla^2 \Phi - 4 (\partial \Phi)^2 - \tfrac{1}{12} H^2 = 0, 
\end{equation} 
which only involves the NS sector\footnote{The fact that this equation does not depend on the RR fields allows us to ignore them. Naturally, other equations couple the NS and RR sectors, however this equation only involves the NS sector. }.  In addition to the earlier assumption that the complementary fluctuations can be found, we will also assume that the original solution has no dilaton, $\Phi = 0$, and no $B$-field, $H=0$, so that (\ref{dil}) is simply $R=0$. With the above Ansatz (\ref{ansatz}), this equation takes the form 
\begin{equation}
\label{orig}
R = -e^{-2A} \left[ 10 (\nabla^2_{AdS_3}  + \nabla^2_{S^3} )A + 20 \partial_{M} A \partial^M A\right] = 0,  
\end{equation}
where the index $M$ ranges over both $AdS_3$ and $S^3$ directions. 

Moving along to the non-Abelian T-dual, we adopt the Ansatz that works when $A$ depends solely on the coordinates of the $AdS_3$ space-time \cite{Itsios:2012dc}, in other words, when it is a singlet, and simply re-introduce dependence on coordinates $(r, \theta, \phi)$ after the T-duality. Note, there is \textit{a priori} no relationship between $(\theta, \phi)$ and any of the original coordinates on the $S^3$. After a plethora of cancellations, one finds that (\ref{dil}) for the non-Abelian T-dual equation simplifies accordingly, 
\begin{eqnarray}
&&-e^{-2A} \left[ 10 (\nabla^2_{AdS_3} A + 20 \partial_{\mu} A \partial^{\mu} A\right] -e^{-2A} \left[ 10 \nabla_{S^2}^2 A + 20 \partial_{\alpha} A \partial^{\alpha} A \right] \\ &&- \frac{e^{2A}}{r^2} \left[ 14 \nabla_{S^2}^2 A + 68 \partial_{\alpha} A \partial^{\alpha} A \right]  - e^{2A} \left[ 14 \partial_r^2 A + 28 \frac{1}{r} \partial_r A + 68 (\partial_r A)^2 \right] = 0, \nonumber
\end{eqnarray}
where $\mu=0,1,2$ and $\alpha=1,2$ denote $AdS_3$ and $S^2$ coordinates, respectively.  

In the first line all $r$-dependence has disappeared and the second Laplacian is confined just to the (unit radius) $S^2$. This closely mirrors the original result (\ref{orig}) and in the strict $r \rightarrow \infty$ limit, all dependence of $A$ on $r$ can be dropped. Naturally, when $A$ is independent of internal directions, we find that the equation is the same before and after T-duality, as expected. 

Now, assuming $A$ is suitably small, we can linearise by dropping quadratic terms and separate the above equation into parts, $A = j(r) Y_{l,m}(\theta, \phi)$, where $Y_{l,m}$ denote standard spherical harmonics. In the process, one encounters the spherical Bessel equation 
\begin{equation}
j'' + \frac{2}{r} j' + \left( k^2 - \frac{l(l+1)}{r^2} \right) = 0,  
\end{equation}
where $k \in \mathbb{R}$ corresponds to the continuous part of the spectrum. Indeed, this is precisely the equation one encounters when one solves the scalar wave equation in a non-Abelian T-dual background \cite{Polychronakos:2010hd} leading to spherical Bessel functions $j_{l} (k r)$, which are regular at the origin where the internal space becomes $\mathbb{R}^3$.  
One expects that the analysis here can be extended to a larger set of fluctuations and that the presence of a warp factor, such as in the $AdS_6$ case, will not affect the conclusion that the spectrum contains both continuous and discrete parts.

\section{Towards a holographic interpretation of the non-Abelian T-dual}

As our non-Abelian dual background is a warped $AdS_6$ geometry, we expect it to be dual to a fixed point theory with $\mathcal{N}=1$ SUSY in 5d, to whose analysis we now turn. Since we have seen that a non compact $r$ direction would lead to a continuous spectrum, we will assume in the following that $r\in [0,R]$ with $R= \pi$. As discussed above, this seems to be the only way to have a globally well-defined SUGRA solution for compact $r$, modulo the (very important) caveat of the termination of the geometry at a regular point. For finite $R$ we might expect four $U(1)$ gauge fields in $AdS_6$. Two would arise from the reduction of 4-form RR potentials over the $S^2$ and either $r$ or $\theta$, and the other two would come from the reduction of the 2-form potential over either $r$ or $\theta$. This would imply a global symmetry group whose Cartan is $U(1)^4$. This fits nicely with the quiver candidate for the dual CFT that we will propose later in this section, even though in the absence of a precise way to impose the cut-off it could well be that less gauge fields existed\footnote{Note that this does not raise an immediate contradiction, since it is common that in backgrounds with backreacted flavor branes the flavor symmetry currents are not apparent in SUGRA.}. On top of this the $SU(2)$ isometry acting on the $S^2$ should correspond to the $SU(2)_R$ R-symmetry of the dual theory.

\subsection{Probing the Coulomb branch}

On general grounds we can probe the Coulomb branch of the theory by considering the supersymmetric locii of probe branes filling $\mathbb{R}_{1,4}$. In the following we examine each such objects separately.

\subsubsection{D5 branes}

A D5-brane wrapped on $\mathbb{R}_{1,4}\times M_1$, where we denote by $M_1$ the space spanned by the $r$ variable, experiences a no-force condition when located at $\theta=0$. This brane should be responsible for the $N_5^\theta$ charge for $n\neq 0$. 

Since this D5-brane does not have indices along the $S^2$ it does not capture the $H_3$ flux. Hence in order to find the corresponding $C_6$ we can just set locally $F_7=dC_6$, finding
\begin{equation}
C_6=-\frac{3^6\,L^6}{2^6}\,r\,\rho^5\,d^5x\wedge dr\, .
\end{equation}
The DBI action reads in turn
\begin{equation}
S_{DBI}=-T_5\,\int e^{-\phi}\,\sqrt{g}=-T_5\,\int \frac{3^6\,L^6\,\rho^5}{2^6}\,\sqrt{r^2+e^{4\,A}}
\end{equation}
Assuming the brane lives at $\theta=0$ this is

\begin{equation}
S_{DBI}=-T_5\,\int \frac{3^6\,L^6}{2^6}\,\rho^5\,r
\end{equation}
which precisely cancels against the WZ term for an anti-D5-brane. Therefore an anti-D5 sitting at $\theta=0$ does not feel a force. 

Upon considering the fluctuations of this brane we find a 5d Chern-Simons term from the WZ action:
\begin{equation}
S_{5d\,CS}=\frac{(2\pi)^3}{6}\,T_5\,\int F_1\,\int A\wedge F\wedge F=-\frac{N_7^r}{24\,\pi^2}\,\int A\wedge F\wedge F\, ,
\end{equation}
with coefficient $N_7^r$. We can also look at the fluctuations of the DBI action to obtain the effective YM coupling. It is easy to see that such fluctuations lead to
\begin{equation}
S=\frac{9\,L^2\,m^{2/3}}{128\, \pi^3}\int \, \cos^{2/3}{\theta}\, \sqrt{r^2+e^{4A}}\, \rho\,F_{\mu\nu}^2
\end{equation}
which, at $\theta=0$, reduce to 

\begin{equation}
S=\frac{9\,L^2\,m^{-1/3}\,N_7^r}{128\,\pi^3}\int \,\rho\,F_{\mu\nu}^2
=\int \frac{1}{g_{D5}^2}\, F_{\mu\nu}^2 \qquad \leadsto \qquad \frac{1}{g_{D5}^2}=\frac{9\,L^2\,m^{-1/3}\,N_7^r}{128\,\pi^3}\,\rho
\end{equation}

We will see in the next subsection that exactly the same theory is obtained by studying the fluctuations of  D7-branes wrapped on  $\mathbb{R}_{1,4}\times M_1\times S^2$, with $N_7^r\leftrightarrow N_5^r$.

Finally let us note that a D5-brane wrapped on $\mathbb{R}_{1,4}$ times the $\theta$ direction does experience a force for any value of $r$, and hence is not supersymmetric.

\subsubsection{D7 branes}

Let us now look at a D7-brane wrapped on $\mathbb{R}_{1,4}\times M_1\times S^2$, which should be responsible for the $N_7^\theta$ charge. This brane now captures the $B_2$ flux.
Taking the conventions $\mathcal{F}=2\pi\,F-B_2$ the CS term reads
\begin{equation}
\label{signWZ}
S_{WZ}=T_7\,\int (C_8-C_6\wedge B_2)\, .
\end{equation}
Using that for our background

\begin{equation}
d\,(C_8-C_6\wedge B_2) = F_9-F_7\wedge B_2 
\end{equation}
we find that

\begin{equation}
C_8-C_6\wedge B_2=\frac{3^6\,L^6}{2^6}\,r^2\,\rho^5\,d^5x\,\wedge dr\wedge {\rm Vol} (S^2)
\end{equation}
and therefore
\begin{equation}
S_{CS}=4\pi\, T_7\,\frac{3^6}{2^6}\, L^6 \int \,r\,(r-\pi\,n)\,\rho^5\,.
\end{equation}
It is easy to check that the DBI action is given by exactly the same expression with opposite sign at $\theta=0$. Therefore, we find another no-force condition when the D7-brane sits at $\theta=0$.
Note that for vanishing $n$ the D7-brane becomes BPS for all $\theta$.

Let us now consider the fluctuations of this brane.  We find from the Chern-Simons action
\begin{equation}
S_{5d\, CS}=T_7\,\frac{(2\pi)^3}{6}\,\int (C_2-C_0\,B )\wedge F\wedge F\wedge F = -T_7\,\frac{(2\pi)^3}{6}\,\int {\hat F}_3\,\int A\wedge F\wedge F 
\end{equation}
where we have used that 
\begin{equation}
d(C_2-C_0\wedge B_2)=F_3-F_1\wedge B_2={\hat F}_3\, ,
\end{equation}
as defined in section 4. Performing the integration we find
\begin{equation}
S_{5d\, CS}=-\frac{N^r_5}{24\,\pi^2}\,\int A\wedge F\wedge F
\end{equation}
that is, a worldvolume Chern-Simons theory with coefficient $N^r_5$. The fluctuations of the DBI action give in turn
\begin{equation}
S=\frac{9\,L^2\,m^{-1/3}\,N^r_5}{128\,\pi^3}\int \,\rho\,F_{\mu\nu}^2
\end{equation}
which is exactly the same expression for the fluctuations of the D5-brane wrapped on $\mathbb{R}_{1,4}\times M_1$, with $N_7^r\leftrightarrow N_5^r$. Note that these are the branes that would have become tensionless for $N_5^r=0$, as discussed in the previous section.

Finally, one can see that D7-branes wrapped on $\mathbb{R}_{1,4}\times S^2$ times the $\theta$ direction and on $AdS_6\times S^2$ do experience a force for any value of $r$ or $(r,\theta)$.

\subsubsection{D7-branes from D5-branes}

In this section we show that the D5 and D7 branes that we have just discussed are related through Myers dielectric effect. We restrict the analysis for simplicity to vanishing large gauge transformations. $n$ in this section will refer to the number of coincident D5-branes.

Schematically the DBI action describing $n$ coincident D5-branes is given by (see \cite{Myers:1999ps}  for more details)
\begin{equation}
S_{DBI}=-T_5 \int {\rm STr} \{ e^{-\phi}\sqrt{g}\sqrt{{\rm det}Q}\}
\end{equation}
where 
\begin{equation}
Q^i_j=\delta^i_j+\frac{i}{2\pi}[X^i,X^k](g-B_2)_{kj}
\end{equation}
and $i,j,k$ run over the transverse non-Abelian directions. Taking the D5-branes to expand into a fuzzy $S^2$ and using Cartesian coordinates we can impose the condition $\sum_{i=1}^3 (x^i)^2=1$ at the level of matrices if the $X^i$ are taken in the irreducible totally symmetric representation of order $n$, with dimension $n+1$,
\begin{equation}
\label{noncom1}
X^i=\frac{1}{\sqrt{n(n+2)}}J^i
\end{equation}
with $J^i$ the generators of $SU(2)$, satisfying $[J^i,J^j]=2i\epsilon_{ijk}J^k$. We then have that
\begin{equation}
\label{noncom2}
[X^i,X^j]=\frac{2i}{\sqrt{n(n+2)}}\epsilon_{ijk} X^k\, .
\end{equation}
Substituting in the DBI action we find a dielectric contribution
\begin{equation}
S_{DBI}=-\frac{T_5}{2\pi}\, \frac{3^6}{2^5}\, L^6\frac{n+1}{\sqrt{n(n+2)}}\int  r^2 \rho^5\, .
\end{equation}
This action gives in the supergravity limit, $n\rightarrow\infty$, the DBI action for a D7-brane wrapped on $\mathbb{R}_{1,4}\times M_1\times S^2$, discussed previously
\begin{equation}
S_{DBI}=-4\pi\, T_7\,\frac{3^6}{2^6}\, L^6 \int \,r^2\rho^5\, .
\end{equation}
For the CS action we find in turn
\begin{equation}
S_{CS}=i\, \frac{T_5}{2\pi}\int (i_X i_X) (C_8-C_6\wedge B_2)=\frac{T_5}{2\pi}\,\frac{3^6}{2^5}\,L^6\frac{n+1}{\sqrt{n(n+2)}}\int  r^2 \rho^5
\end{equation}
We then see that for a system of coincident D5-branes the monopole couplings, dominant in the supergravity limit, cancel at $\theta=0$, giving rise to a no-force condition. The dipole couplings, in turn, give in the large $n$ limit the action describing a D7-brane wrapped on $\mathbb{R}_{1,4}\times M_1\times S^2$, which in the absence of large gauge transformations is supersymmetric for all $\theta$.

\subsection{Instantons}

In the previous subsection we have seen that the Coulomb branch of our putative dual CFT seems to be two dimensional, as we have two branes --the D5 extended on $r$ and the $D7$ extended on $r$ and $S^2$-- which we can move independently. This would naively suggest a dual theory with two gauge groups. On the other hand, in five dimensions each vector multiplet automatically comes with a topologically conserved instantonic current. Hence, our naive identification would demand two types of instantonic particles. Note that on the Coulomb branch these must be non-gauge invariant, being the charge proportional to the Chern-Simons term of the corresponding $U(1)$ Coulomb branch. In the gravitational dual this translates into the fact that these instanton states must be dual to wrapped branes with a tadpole given by the CS coefficients, which we found to be $N_5^r$ and $N_7^r$.

\subsubsection{D1 instantons}

Let us consider a D1-brane wrapping the $M_1$ space. This brane has the expected world volume  tadpole coming from the WZ coupling:
\begin{equation}
S_{WZ}=-2\pi\,T_1\,\int F_1\,\int A_t=-N_7^r\,\int A_t 
\end{equation}
Its DBI action reads in turn
\begin{equation}
S_{DBI}= -\frac94\, T_1\, L^2\, m^{2/3} \cos^{2/3}{\theta}\int\sqrt{r^2+e^{4A}}\, \rho
\end{equation}
and therefore vanishes at $\theta=\frac{\pi}{2}$, while we find for $\theta=0$:
\begin{equation}
S=-\int \frac{16\,\pi^2}{g_{D5}^2}
\end{equation}
which is the expectation for an instantonic particle.

For multiple D1-branes we can consider as well the dielectric couplings
\begin{equation}
S_{WZ}^{\rm diel}=-i \,T_1\int (i_X i_X)(F_3-F_1\wedge B_2)\wedge A=-i\, T_1\int (i_X i_X)\hat{F}_3\wedge A
\end{equation}
These terms are responsible for the expansion of the D1-branes into a D3-brane wrapped on the internal $S^2$. Taking 
 ${\hat F}_3$ as given in (\ref{hatsfields}) \footnote{As in the previous subsection we restrict the analysis to zero large gauge transformations.} and the non-commutative ansatz given by (\ref{noncom1}) we find for a set of $n$ D1-branes:
 \begin{equation}
 \label{D1diel}
S_{WZ}^{\rm diel}=-\frac{n+1}{\sqrt{n(n+2)}}\, N_5^r\int  A_t
\end{equation}
The action (\ref{D1diel}) gives in the large $n$ limit the WZ action of a D3-brane
wrapped on $M_1\times S^2$, which, as we show in the next subsection, has a tadpole with charge equal to $N_5^r$. The DBI action 
\begin{equation}
S_{DBI}=-T_1 \int {\rm STr} \{ e^{-\phi}\sqrt{g}\sqrt{{\rm det}Q}\}\supset S_{DBI}^{\rm diel}
\end{equation}
where, as in the previous section 
\begin{equation}
Q^i_j=\delta^i_j+\frac{i}{2\pi}[X^i,X^k](g-B_2)_{kj}\, ,
\end{equation}
gives in turn the following dielectric contribution
\begin{equation}
S_{DBI}^{\rm diel}
=-T_1\, \frac{9}{4\pi}\,L^2\, m^{2/3}\cos^{2/3}{\theta}\,\frac{n+1}{\sqrt{n(n+2)}}\int \rho\, r^2\, .
\end{equation}
Taking $\theta=0$ this gives
\begin{equation}
\label{DBIdiel}
S_{DBI}^{\rm diel}=
-\frac{n+1}{\sqrt{n(n+2)}}\int \frac{16\pi^2}{g_{D7}^2}
\end{equation}
which reproduces in the large $n$ limit the action for a D3-brane wrapped on $M_1\times S^2$.

\subsubsection{D3 instantons}

To complete the analysis of the previous subsection let us now consider a D3-brane wrapping the $M_1\times S^2$ space. This brane has a tadpole
\begin{equation}
S_{WZ}=-2\pi\,T_3 \int {\hat F}_3\,\int A_t
\end{equation}
with ${\hat F}_3$ given in (\ref{hatsfields}). Integrating over the $S^2$ we find the expected tadpole
\begin{equation}
S_{WZ}=-\,N_5^r\,\int A_t\, .
\end{equation}
The DBI action reads in turn
\begin{equation}
S_{DBI}=-9\pi\, T_3\, L^2\,m^{2/3}\,\cos^{2/3}\theta\,\int \rho\,r^2
\end{equation}
As before, this vanishes for $\theta=\frac{\pi}{2}$, while we find for $\theta=0$:
\begin{equation}
S=-\int \frac{16\,\pi^2}{g_{D7}^2}
\end{equation}
which is the expectation for an instantonic particle.

\subsection{Flavors and D5-branes wrapping $AdS_6$}

We can find a BPS D5-brane wrapped on the $AdS_6$ spacetime and located at $\theta=\pi/2$, $r=0$, which should  be responsible for the charges $N_7^r$ and $N_5^r$, which as we have seen are not independent. Indeed, being extended along the infinite $AdS$ radial direction, this brane cannot be interpreted as a color brane. Instead, it should correspond to a global flavor symmetry in the dual theory. 

The relevant RR-potential reads
\begin{equation}
C_6=-\frac{3^6}{2^7}L^6\Bigl[5r^2+\frac{3}{2^3}W^4 L^4(1+\frac12 \cos^2{\theta})\Bigr]\rho^4d\rho\wedge d^5 x
\end{equation}
The DBI action reads in turn
\begin{equation}
S_{DBI}=-T_5\,\int e^{-\phi}\,\sqrt{g}=-T_5\,\frac{3^7}{2^8}\int\, W^2 L^8 \sin{\theta}\, \rho^4 \sqrt{r^2+e^{4A}}
\end{equation}
Assuming the brane lives at $r=0$ this is
\begin{equation}
S_{DBI}=-T_5\, \frac{3^7}{2^{10}}\int\, W^4 L^{10} \sin^3{\theta}\,\rho^4
\end{equation}
which precisely cancels the CS term at $\theta=\pi/2$. Thus the D5-brane experiences a no-force condition precisely when located at what would be the naive location of the orientifold fixed plane in the dual background (see the Discussion).

Starting from multiple D5-branes it is easy to see that they can expand into D7-branes wrapped on $AdS_6\times S^2$, which become however non-BPS for any value of $r$, $\theta$.  The relevant dielectric terms read\footnote{Again in this calculation we set to zero the large gauge transformations of $B_2$ and $n$ refers to the number of coincident branes.}
\begin{equation}
\label{D7diel}
S=-\frac{T_5}{2\pi}\, \frac{3^7}{2^7}L^8 \frac{n+1}{\sqrt{n(n+2)}}\int W^2 \sin{\theta}\, r^2\, \rho^4
+\frac{T_5}{2\pi}\,\frac{3^5\, 5}{2^5}L^6 \frac{n+1}{\sqrt{n(n+2)}}\int r^3 \rho^4\, ,
\end{equation}
which give in the large $n$ limit the action for a D7-brane wrapped on $AdS_6\times S^2$. As we can see from (\ref{D7diel}) this brane experiences a force unless $r=0$, in which case each term identically vanishes.

\subsection{A dual CFT with two gauge groups? Ranks and branes with tadpoles}

In the preceding subsections we have seen that the probe brane analysis of the background is consistent with a putative dual CFT with two gauge groups, with induced CS levels $N_5^r$, $N_7^r$.
This suggests that the gauge groups see a number of flavors proportional to $N_5^r$ and $N_7^r$, respectively. 

In order to elucidate the rank of the gauge groups we now turn to the analysis of baryon-like operators, since on general grounds these should be dual to branes wrapped in the internal geometry with a tadpole that is proportional to the rank of the gauge group.

In the original $AdS_6$ background a D4-brane wrapped on the $\theta$ direction times the $S^3$ in the internal space develops a tadpole with charge $N$, the number of color D4-branes.
This, would be, baryon vertex is however removed from the spectrum by the orbifold projection
$I_\theta: \theta\rightarrow\pi -\theta$. This corresponds to the fact that $USp$ baryons are unstable against decay into mesons. Given that in the non-Abelian dual background 
global properties --in particular orientifold projections-- are unclear\footnote{See however the Discussion.}, it is not obvious whether similar baryons will actually be stable or not. Nevertheless, blindly considering them will give us qualitative information about the rank of the dual gauge groups, as it did in the original $AdS_6$ geometry.

On one hand, a D1-brane wrapped on the $\theta$ direction has a tadpole with charge $N_7^\theta$ coming from the $F_1$ flux
\begin{equation}
S_{WZ}=2\pi\,T_1\,\int d\theta \, G_1 \, \int A_t=-N^{\theta}_7\,\int A_t
\end{equation}

On the other hand, a D3-brane wrapped on $\theta$ and the $S^2$ captures the $\theta$ component of the 3-form flux, inducing a tadpole of
\begin{equation}
S_{WZ}=2n \pi^2\,T_3\,{\rm Vol}(S^2) \int d\theta \, G_1 \, \int A_t=-n N^{\theta}_7\,\int A_t=
-N^{\theta}_5\,\int A_t
\end{equation}

Thus, we seem to find two baryon vertices, consistent with the two gauge groups which we have conjectured. Furthermore, the tadpoles suggest that the ranks of the gauge groups are proportional to $N_7^{\theta}$ and $N_5^{\theta}$. Note that the latter is proportional to $n$ through eq. (\ref{N5}), which is actually vanishing in our background with $r\leq R$, $R=\pi$. It is tempting to speculate that this might be related to the origin of the subtle behaviour of the background.

\subsection{Putting it all together: a conjecture for the dual CFT}

\begin{figure}[h!]
\centering
\includegraphics[scale=.4]{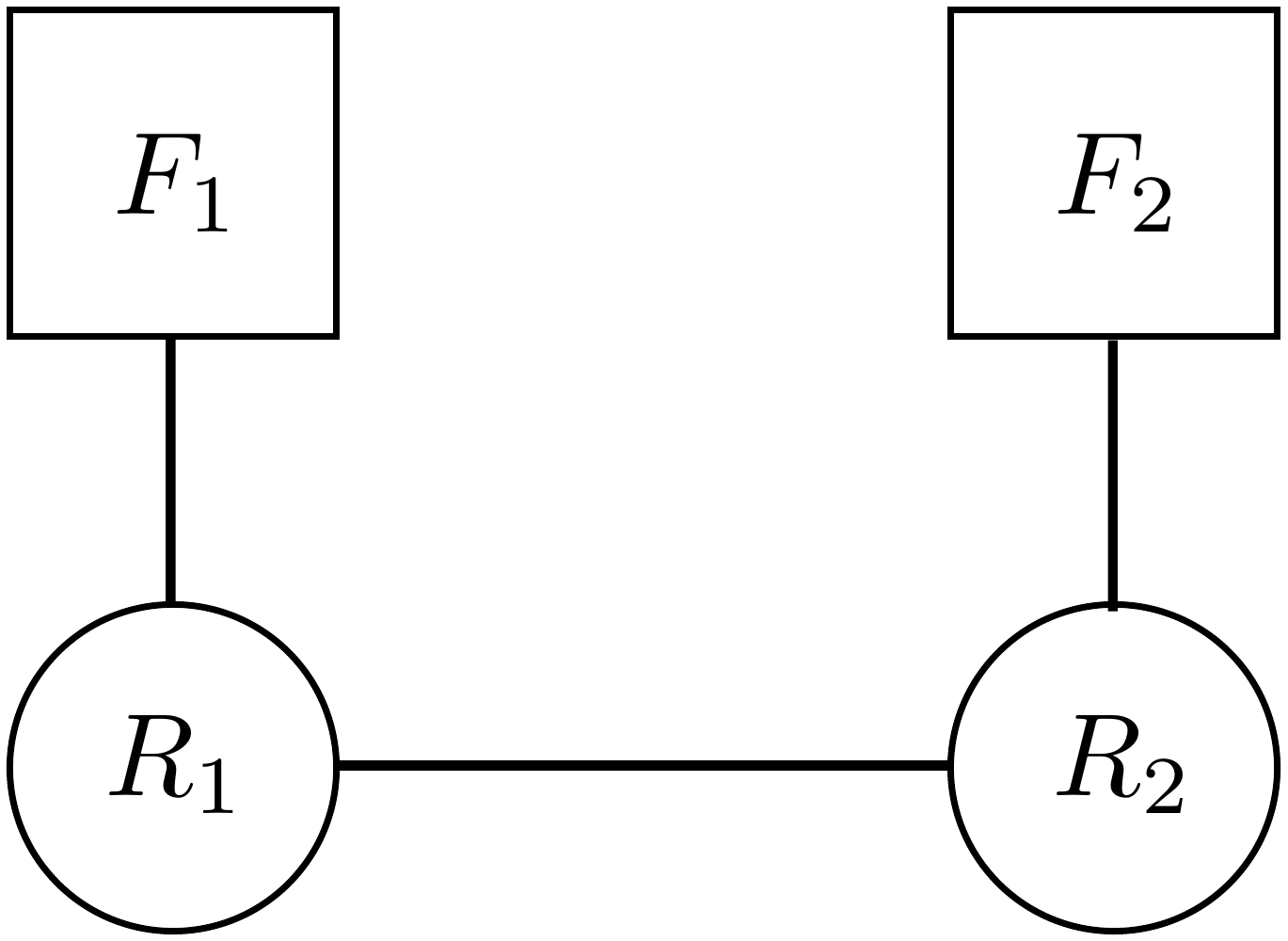}
\caption{\label{theory} Schematic proposal for the dual CFT.}
\end{figure}

The presence of two directions in the Coulomb branch, together with the existence of two instantonic particles, suggest a dual CFT with two gauge groups. Let us call $R_1$ and $R_2$ their corresponding ranks. There are also non-compact branes, which should correspond to flavor symmetries. Since on general grounds we expect one flavor symmetry, $F_i$, for each gauge group, a schematic proposal for the dual CFT could be as shown in Fig. \ref{theory}. Moreover,
identifying $R_1=N_7^\theta$, $R_2=N_5^\theta$ we have that each gauge group should feel, respectively, $F_1+R_2=N_5^r$ and $F_2+R_1=N_7^r$ flavors.

Note that as shown in section 4, $N_5^\theta$ should actually be zero in order to have a globally well-defined dual background for compact $r$. Therefore, strictly speaking we find in that case a, to say the least, subtle dual CFT, since it contains a fully depleted gauge group. It is tempting to conjecture that this
happens in the CFT as a consequence of the fact that 
 we had to terminate the background at a point that is perfectly regular, in order to find well-defined quantization conditions. It may be that a clear prescription for the global properties of the dual background could generate a perfectly regular background for arbitrary large gauge transformations with non-depleted gauge groups, such that the non-Abelian T-dual geometry arises as a limit of this conjectured background.
 
The fate of global orientifold-like identifications in the dual theory is another global aspect that 
cannot be worked out with our current knowledge of non-Abelian T-duality. Although in general grounds we expect the dual CFT to involve two baryonic $U(1)$ symmetries and two topological $U(1)_T$ symmetries, giving rise to the expected four $U(1)'s$, it is not clear whether the baryon vertices would remain in the spectrum after the dual orientifold projections. In the original background the orientifold projection was indeed forbidding the baryon vertex, consistently with the fact that there are no $Sp$ baryons. Therefore we cannot elucidate whether the gauge groups are unitary, symplectic or orthogonal --or some more exotic possibilities--.

\subsection{On the Higgs branch}

The dual CFT that we have just proposed has been designed to provide quantitative agreement between the Coulomb branch and the spectrum of branes extended in $\mathbb{R}_{1,4}$. 
If this proposal is to be sensible we should also expect a whole Higgs branch in the moduli space where hypermultiplets take VEVs.

The situation in the original $AdS_6$ background is such that the 5d dual CFT contains operators in the Higgs branch which correspond in the gravity dual to giant gravitons sitting on top of the O8/D8 system \cite{Bergman:2012qh}. In particular, mesonic operators made out of bifundamental or antisymmetric fields correspond to D4-brane dual giant gravitons \cite{Grisaru:2000zn, Hashimoto:2000zp} wrapped on an $S^4$ submanifold inside $AdS_6$ and propagating on the fiber and/or the azimuthal angle of the internal $S^3$. Note that these configurations only capture the part of the Higgs branch not involving fundamental fields.  

The natural candidates for similar giant graviton configurations in the non-Abelian dual background are D5 (D7) branes wrapped on $S^4 \times M_1$
($S^4 \times M_1\times S^2$), with $S^4\subset AdS_6$. In this case these branes can only propagate on the azimuthal angle of the $S^2$, given that the fiber direction of the $S^3$ disappears after the dualization. Note that this already poses a problem, as the phase space corresponding to such branes would not be of complex dimension 2, and hence cannot be a hyperk\"ahler variety as one would expect for a Higgs branch. Nevertheless one can explicitly show that indeed these candidates do not behave as giant gravitons. 

Let us  consider first a D5-brane moving on $\psi$, the azimuthal angle of the internal $S^2$, and wrapped on $S^4\times M_1$, with $S^4\subset AdS_6$. 
We consider global coordinates for the $AdS_6$ part of the geometry:
\begin{equation}
ds^2(AdS_6)=-(1+\rho^2)\,dt^2+\frac{1}{(1+\rho^2)}\,d\rho^2+\rho^2\,ds^2(S^4)\, , \qquad d{\rm Vol}(AdS_6)=\rho^4\,dt\wedge d\rho\wedge d{\rm Vol}(S^4)
\end{equation}
This brane couples to the RR potential
\begin{equation}
C_6=\frac{3^6}{2^6}\,L^6r\,\rho^5\,dt\wedge {\rm Vol}(S^4) \wedge dr\, ,
\end{equation}
with the full action given by
\begin{eqnarray}
S=&&-\frac{3^6}{2^6}\,T_5\, {\rm Vol}(S^4)\,L^6\,\int \, \rho^4\, \sqrt{r^2+e^{4\,A}}\, \sqrt{1+\rho^2-\frac{r^2\,\sin^2{\theta}\,\sin^2\xi}{9\,(r^2+e^{4\,A})}\,\dot{\psi}^2}+\nonumber\\
&&+\frac{3^6}{2^6}\,T_5\,{\rm Vol}(S^4)\,L^6\,\int\,r\,\rho^5
\end{eqnarray}
The equation of motion for $\xi$ is satisfied for $\xi=0$ and $\pi/2$. Obviously only the latter can be a giant. However the equation of motion for $\theta$ gives $\theta=0$ as the only solution, and here the D5-brane does not propagate.

A similar calculation for a D7-brane wrapped on $S^4\times M_1\times S^2$ and propagating on the, now worldvolume, $\psi$ direction gives
\begin{equation}
S=-\frac{3^6}{2^6}\,T_7\, {\rm Vol}(S^4)\,L^6\, {\rm Vol}(S^2)\Bigl[ \int \, \rho^4\, r^2 \sqrt{1+\rho^2}
-r^2\,\rho^5\Bigr]
\end{equation}
where we can see that the dependence on $\dot{\psi}$ simply disappears for any $\theta$.

Thus, the gravity dual suggests that the dual CFT has no Higgs branch. Our proposed gauge theory seems to have however a Higgs branch, with operators involving fundamental fields as well as operators made only of  bifundamentals. That the former are not seen in the non-Abelian dual background could be expected given that already in the original $AdS_6$ background the D4 dual giant gravitons were only capturing the subset of the Higgs branch not involving fundamental fields (\textit{i.e.} those associated to $D8-D4$ strings)  \cite{Bergman:2012qh}. In other words, the $SO(2\,N_f)$ is not visible in the geometry. However, we would expect the operators associated to mesons made out of bifundamentals to span a Higgs branch captured by the gravity dual, which is however not present. Since for compact $r$ one of the gauge groups has zero rank one could argue that these 
fields are not really present and explain in this way the apparent non-existence of a Higgs branch in the dual geometry.

\section{Entanglement entropy}

The entanglement entropy \cite{Ryu:2006bv,Klebanov:2007ws} can be determined through calculating \cite{Jafferis:2012iv}
\begin{equation}
S = \frac{4\pi}{2\kappa_{10}^2}\int d^8x\, e^{-2 \phi} \sqrt{g}, 
\end{equation}
where $g$ is the induced eight-dimensional metric in string frame. The result of this calculation should provide information about the $S^5$ free energy of the dual CFT, following \cite{Casini:2011kv}.

The measure $e^{-2 \phi} \sqrt{g}$ is a well-known invariant of (Abelian) T-duality, so we could expect the entanglement entropy of the non-Abelian T-dual geometry to also resemble the first. Following a similar calculation in \cite{Jafferis:2012iv}, the difference we can quantify by comparing the entropy of the original background (\ref{IIAbackground}) against that of its non-Abelian T-dual (\ref{NATD}) 

\begin{equation}
S = \frac{3^7 L^{10} R^3 m^{\frac{1}{3}} \pi^4}{2^{6}\, 5 }\frac{1}{2{\tilde \kappa}_{10}^2} \int \frac{\rho(z)^3 \sqrt{1 + \rho'(z)^2}}{z^4} dz 
\end{equation} 
where we have allowed for a different $\kappa_{10}$ as required by the correct quantization of the charges. The minimal surface equation is solved for
\begin{equation}
\rho=\sqrt{{\cal R}^2-z^2}.
\end{equation}
When the integral with respect to $z$ is performed and the universal part extracted, we get an additional factor of $\frac{2}{3}$, leading to a free energy in terms of the conserved charges $N_7^\theta$, $N_7^r$
\begin{equation}
F=-(2{\tilde \kappa}_{10}^2) \frac{3R^4}{5\, 2^7 \pi^{10}} \frac{(N_7^\theta)^{5/2}}{(N_7^r)^{1/2}}\, .
\end{equation}
The free energy in the original theory reads in turn \cite{Jafferis:2012iv}
\begin{equation}
F_{{\rm orig}}=-\frac{9 \pi^{1/2}}{5}\frac{N_4^{5/2}}{m^{1/2}}\, .
\end{equation}

One can then see that for ${\tilde \kappa}_{10}^2=(2\pi)^8\, q$, as required by the correct quantizations of the charges of the dual background, the free energies satisfy
\begin{equation}
F=\frac{\pi}{2}\, p\, F_{{\rm orig}}
\end{equation}
where $p$ is the integer satisfying $N_5^r=2\pi p\, m$. Thus, the $S^5$ free energies of the dual CFT's, differ by a constant in the original and non-Abelian dual backgrounds\footnote{We remark that if the entanglement entropies were to agree, then the six-dimensional Newton constants, as evaluated in \cite{Figueras:2013jja}, would also be the same. In this case however the charges are not properly quantized in the dual background.}.

\section{Discussion}

An important drawback of non-Abelian T-duality is that it cannot be used to extract global properties of the dual space. This global information is however crucial in order to find out the string theory realization of the dual solution. Combining partial information derived from non-Abelian T-duality with consistency requirements on the dual CFT we have proposed a candidate dual CFT with two gauge groups. Some of these consistency requirements have in turn been used  to extract partial global information about the newly generated geometry. Indeed, 
our construction works on the basis of a compact $M_1\times S^2$ dual space. This poses on the other hand an important puzzle on the dual theory, with the geometry terminating at a perfectly well-defined point where no invariant quantity blows up. As raised above, it is tempting to conjecture that the termination of the geometry at a smooth point is intimately related to the depletion of the rank of one of the gauge groups. It might well be that there is a regular solution which can be extended beyond $R=\pi$ and which does not have a depleted gauge group. Note that such depletion is very reminiscent of the cascade, which, for theories with 8 supercharges, is more accurately thought of as a Higgsing sequence \cite{Aharony:2000pp}. Thus it might well be that there exists a regular and well-defined solution such that, upon appropriately choosing a point on its moduli space, the non-Abelian T-dual background is the effective description at least in some range of the coordinates. Note that in the end we are performing a (non-Abelian) T-duality transformation, which naively would result in smeared brane configurations hence somehow choosing a point in the Coulomb branch; intuition which is also reminiscent of these lines. The lack of global information through non-Abelian duality does not allow however to explore further this  and other open issues left out by our analysis. 

One such open issue is the nature of the dual gauge groups. Finding this out requires precise knowledge of the orientifold projection in the dual theory.
The picture that seems to arise is that a D$p$-brane in the original background wrapped on the $S^3$ that is being dualized gives rise to both a D$(p-1)$-brane transverse to the $M_1$ and a D$(p-3)$-brane transverse to $M_1\times S^2$. If on the other hand the original D$p$-brane is transverse to the $S^3$, both a D$(p+1)$-brane wrapped on $M_1$ and a D$(p+3)$-brane wrapped on $M_1\times S^2$ arise. Showing that the two dual branes coming out from the same original brane carry charges that are not independent requires however some non-trivial input.

We have seen through our analysis that the D5 and D7 charges that arise from the D8-brane of the original $AdS_6$ background depend on the cut-off that must be imposed on the $M_1$ space such that a continuous spectrum of fluctuations can be avoided. The way the value of this cut-off is set is however quite subtle, coming from imposing global consistency on the dual background under large gauge transformations of the $B$-field. This consistency requirement allows to also  set to zero the charge of the D5-branes wrapped on $M_1$ dual to the D4-branes of the original background, leaving the D7-branes wrapped on $M_1\times S^2$ as the only color branes in the dual theory. We would like to stress that even if non-contractible 2-cycles turn out not to exist in the dual geometry for finite $r$, these branes would still be supersymmetric and their stability would therefore be guaranteed.

Within the previous ``phenomenological'' picture, the $O8^-$ fixed plane of the original background would be mapped under the non-Abelian T-duality onto a $O7^-$ fixed plane transverse to $M_1$ and a $O5^-$ fixed plane transverse to $M_1\times S^2$. The mapping of the $I_\theta \Omega$ orientifold action of the original geometry under the transformation: 
$g^{-1} \partial_+ g=M^T \partial_+ \chi$, $g^{-1} \partial_- g=-M \partial_- \chi$, responsible at the level of the sigma-model of the non-Abelian T-dual background (see \cite{Lozano:1995jx}) gives a dual $I_\theta I_\chi \Omega$ orientifold action, which suggests that the $O8^-$ is mapped onto both a $O5^-$ orientifold fixed plane located at $\theta=\pi/2$, $r=0$ and a $O7^-$ plane located at the same place but wrapped on the $S^2$. The second turns out however to be non-BPS through a similar analysis as that performed in section 5.4. The D5-branes wrapped on $AdS_6$ discussed in section 5.4 are however BPS exactly at the location of the $O5^-$ orientifold fixed plane. It would be interesting to elucidate whether the dual background that we have constructed came out indeed as the near horizon geometry of this D5-D7 system, thus realizing the conjecture suggested above. 

The picture of 4 gauge fields in AdS seems to fit nicely in the quiver that we have proposed  --a flavored 2-node quiver in 5d would come with 2 baryonic currents and 2 instantonic currents captured by RR potentials--. However the existence of such gauge fields in $AdS$ does depend on global issues of the background which are not well under control. We would like to stress that even if some of these gauge fields turned out not to exist this would not raise an immediate contradiction. In fact, it would not be surprising that symmetries associated to backreacted flavor branes are not seen --\textit{e.g.} in the original Brandhuber-Oz solution the $SO(2\,N_f)$ currents are just not seen in the $AdS_6$ background--.

As we have seen, supersymmetric probes corresponding to operators in any putative dual CFT are not guaranteed to survive the non-Abelian T-duality process. A recent example that springs to mind is the existence of a supersymmetric M2-brane probe in the context of Ref. \cite{Itsios:2013wd}, which should correspond to the canonical BPS operator identified in \cite{Gaiotto:2009gz}. Interestingly, there is a well-known theorem \cite{Gauntlett:2006ai} identifying supersymmetric embeddings of M2-branes dual to chiral primary operators in the most general class of such geometries \cite{Gauntlett:2004zh}, and it would be instructive to reconcile these results. 

Lastly, in the light of the uniqueness statement for supersymmetric solutions in massive IIA \cite{Passias:2012vp}, and recent success in the identification of (numerical) supersymmetric $AdS_7$ solutions in massive IIA \cite{Apruzzi:2013yva}\footnote{The same paper shows that there are no solutions in Type IIB.}, it is an open direction to classify the supersymmetric $AdS_6$ solutions to Type IIB supergravity in the hope that a future classification may reveal new solutions of relevance to AdS/CFT. Alternatively, if none exist, a statement confining solutions to the Abelian and non-Abelian T-dual of the Brandhuber-Oz solution \cite{Brandhuber:1999np} would be welcome. 

\section*{Acknowledgements} 
We would like to thank Carlos Nu\~nez, Alfonso Ramallo and Kostas Sfetsos for very useful discussions. The authors are partially supported by the Spanish Ministry of Science and Education grant FPA2012-35043-C02-02. D.R-G is also partially supported by the Ram\'on y Cajal fellowship RyC-2011-07593. 

\appendix 

\section{Hopf T-duality}
For completeness in this section we illustrate Hopf T-duality for the original $AdS_6 \times S^4$ space-time of massive IIA supergravity \cite{Brandhuber:1999np}. To the extent of our knowledge, Hopf  T-duality of $AdS_6 \times S^4$ first appeared in \cite{Cvetic:2000cj}, however the implications for supersymmetry post T-duality were not discussed. Notable examples of Hopf T-duality in the literature include \cite{Duff:1998us} and \cite{Duff:1998cr}, where in the case of the former, supersymmetry is broken. 

Here we will confirm that Hopf T-duality on the Brandhuber-Oz solution \cite{Brandhuber:1999np} preserves all supersymmetry at the level of supergravity. This observation is very much in line with expectations, since once the $S^3$ corresponding to the $SO(4)$ isometry is written in terms of a Hopf-fibration, the manifest symmetry becomes $U(1) \times SU(2)_R$, where $U(1)$ is a global symmetry and the Killing spinors do not depend on this direction. The explicit form of the original Killing spinor can be found in (\ref{mIIA_KS}). 
 
Now, performing the T-duality in the standard way, the Hopf T-dual solution takes the form 
\begin{eqnarray}
ds^2 &=& \frac{1}{4} W^2 L^2 \left[ 9 ds^2(AdS_6) + 4 d \theta^2 +  \sin^2 \theta \left( d \phi_1^2 + \sin^2 \phi_1 d \phi_2^2\right)  + \frac{16}{W^4 L^4 \sin^2 \theta} d \phi_3^2  \right], \nonumber \\
B &=& - \cos \phi_1 d \phi_2 \wedge d \phi_3, \quad
e^{\phi} = \frac{4 }{3 L^2 (m \cos \theta)^{2/3} \sin \theta}, \nonumber \\ 
F_3 &=&  \frac{5}{8} L^{4} (m \cos \theta)^{1/3} \sin^3 \theta \sin \phi_1 d \theta \wedge d \phi_1 \wedge d \phi_2, \quad F_1 =  m  d \phi_3.  
\end{eqnarray}

This satisfies the equations of motion, so one just needs to check supersymmetry. Borrowing our conventions from \cite{Itsios:2012dc,hassan}, we can plug this solution into the Killing spinor equations. The dilatino variation 
implies that the underlying projection condition is 
\begin{equation}
\label{hopfproj}
\left[\cos \theta \Gamma^{\phi_1 \phi_2 \phi_3 \theta} \sigma^3 + \sin \theta \Gamma^{\phi_3 \theta} i \sigma^2 \right] \eta =  \eta. 
\end{equation}
Now if we momentarily add tildes to the above Killing spinors, we can compare our new projector (\ref{hopfproj}) with the original projector (\ref{ads6s4proj1})
\begin{equation}
\tilde{\epsilon}_+ = \epsilon_+, \quad \tilde{\epsilon}_- = \Gamma^{\phi_3} \epsilon_-, 
\end{equation}
where we have decomposed the Killing spinors $\eta, \tilde{\eta}$ in terms of their Majorana-Weyl component spinors, $\epsilon_{\pm}, \tilde{\epsilon}_{\pm}$. Note here that the presence of $\Gamma^{\phi_3}$ means that the chirality of $\epsilon_-$ is flipped, which is expected in the transition from massive IIA to type IIB supergravity.

Using our single projection condition, the remaining conditions from the vanishing of the gravitino variations may be solved in turn leading to the solution 
\begin{equation}
\eta = (\cos \theta)^{-1/12} e^{- \frac{\theta}{2}\Gamma^{\phi_1 \phi_2} \sigma^1} e^{- \frac{\phi_1}{2}\Gamma^{\phi_1 \theta} } e^{- \frac{\phi_2}{2}\Gamma^{\phi_2 \phi_1} }  \tilde{\eta}, 
\end{equation}
where $\tilde{\eta}$ is a solution to the Killing spinor equation on $AdS_6$, $\nabla_{\mu} \tilde{\eta} = \frac{1}{2} \Gamma_{\mu} \Gamma^{\theta \phi_1 \phi_2} \sigma^1 \tilde{\eta}$. This concludes our illustration of the preserved supersymmetry of the Hopf T-dual. 

\section{Supersymmetry of the non-Abelian T-dual}

In this Appendix we follow arguments presented in \cite{Itsios:2012dc} and demonstrate that the effect of an $SU(2)$ transformation for space-times with $SO(4)$ isometry\footnote{See section 4 of  \cite{Itsios:2012dc} for further details of the transformation from massive IIA to Type IIB supergravity.} is simply a rotation on the Killing spinors. The calculations presented here generalise the analysis of \cite{Itsios:2012dc} to include transformations from massive IIA to Type IIB supergravity and provide details necessary to support statements in \cite{Lozano:2012au}.

Once again the key observation will be that there is a rotation of the Killing spinor \cite{Itsios:2012dc}
\begin{equation}
\label{rotate}
\eta =  e^{X} \tilde{\eta} =  \exp \left( - \frac{1}{2} \tan^{-1} \left( \frac{e^{2A}}{r} \right)
\Gamma^{\alpha_1 \alpha_2} \sigma^3 \right) \tilde{\eta}\ , 
\end{equation}
where $A$ is an overall warp factor for the $S^3$ of the original space-time, $r$ is the T-dual coordinate in $[0,R]$ and $\eta, \tilde{\eta}$ are Killing spinors for the T-dual and original geometry respectively. In addition, to avoid confusion with the $\theta$ direction we have introduced the coordinates $\alpha_i, ~i=1,2$ to parametrise the residual two-sphere that encodes the $SU(2)$ R-symmetry. 

Once this rotation is taken into account, it is a straightforward exercise to see how the Killing spinor equations for the T-dual geometry can be recast in terms of the Killing spinor equations of the original geometry. 
We begin by examining the gravitino variation in the $r$ direction. After rearranging appropriately, this takes the form 
\begin{eqnarray}
\label{psir}
&&\delta \psi_{r} = e^{X} \biggl[ \frac{1}{2} \slashed{\partial} A \Gamma_{r} - \frac{e^{-A}}{4} \Gamma^{\alpha_1 \alpha_2} \sigma^3 + \frac{e^{\phi}}{8} \biggl(  m i \sigma^2  \\ && \phantom{xxxxxxxxxxxxx} +e^{-3 A} \slashed{G}_1 \Gamma^{r \alpha_1 \alpha_2} \sigma^1  + \slashed{G}_2 \sigma^1  - \slashed{G}_3 \Gamma^{r \alpha_1 \alpha_2} i \sigma^2 \biggr) \biggr] \tilde{\eta} = 0, \nonumber
\end{eqnarray}
where we have redefined $G_3 = *_7 G_4$. 

The strategy now is to show that all the remaining Type IIB Killing spinor equations can be expressed in terms of the original IIA Killing spinor equations and the variation $\delta \psi_r$. As an immediate consequence, when $\delta \psi_r$ is set to zero, we will be able to identify all the conditions on the Killing spinors of the T-dual background. 

Once the Killing spinors are rotated as prescribed by (\ref{rotate}), the gravitino variation along the directions  unaffected by the duality transformation becomes 
\begin{eqnarray}
\label{susy2}
&& \delta \psi_{\mu} = e^{X} \biggl[ \nabla_{\mu}  - \frac{1}{8} H_{\mu \nu \rho}
\Gamma^{\nu \rho} \sigma^3  + \frac{e^{\Phi}}{8} \biggl(m \Gamma^{r} i \sigma^2 \nonumber \\ && \phantom{xxxxxxxxxxx} + 
e^{-3 A} \slashed{G}_1 \Gamma^{\alpha_1 \alpha_2 }  \sigma^1 + \slashed{G}_2 \Gamma^{r} \sigma^1  - \slashed{G}_3 \Gamma^{\alpha_1 \alpha_2} i \sigma^2 \biggr)  \Gamma_{\mu} \biggr] \tilde{\eta}\ .
\end{eqnarray}
This IIB variation can be mapped back to the corresponding Killing spinor equation for Type IIA by employing the redefinitions: 
\begin{equation}
\label{redefine}
\tilde{\epsilon}_+ = \Gamma^{7} \epsilon_+, \quad \tilde{\epsilon}_- = \epsilon_-, \quad \Gamma^{r \alpha_1 \alpha_2} = - \Gamma^{789}. 
\end{equation}
So we can conclude that the gravitino variations in these directions  are satisfied provided the original geometry preserved some supersymmetry. 

We now focus on the residual $S^2$ corresponding to the R symmetry. For concreteness, we analyse only one of the directions on the $S^2$ with the other following from symmetry. The gravitino variation along the $\alpha_1$ direction may be written as 
\begin{eqnarray}
&& \delta \psi_{\alpha_1} = \frac{e^{2 X}}{\sqrt{r^2 + e^{4 A}}} \left( e^{2A} \Gamma^{r \alpha_2} \sigma^3
 - \frac{e^{4A}}{r} \Gamma^{r \alpha_1} \right) \delta \psi_r
\nonumber\\
&& \phantom{xx}
+ \frac{e^X \sqrt{r^2+e^{4A}}}{r}  \biggl[ e^{-A} \partial_{\alpha_1} + \frac{1}{2} \Gamma^{\alpha_1 \mu}
\partial_{\mu} A + \frac{e^{-A}}{4} \Gamma^{r \alpha_2} \sigma^3 + \frac{e^{\Phi}}{8}
\biggl( m \Gamma^{r \alpha_1} i \sigma^2 \nonumber \\
 &&- e^{-3A} \slashed{G}_1 \Gamma^{\alpha_2} \sigma^1 + \slashed{G}_2 \Gamma^{r \alpha_1} \sigma^1  + \slashed{G}_3 \Gamma^{ \alpha_2} i \sigma^2\biggr) \biggr] \tilde{\eta}\ . 
\end{eqnarray}
Using the expression for $\delta \psi_r$ again, we can bring this equation to the simpler form 
\begin{eqnarray}
\delta \psi_{\alpha_1} = e^{4 X} \Gamma^{r \alpha_1} \delta \psi_{r}
 +  \frac{e^{-A} \sqrt{r^2+e^{4A}}}{r} \left(  \partial_{\alpha_1}
  + \frac{1}{2} \Gamma^{r \alpha_2} \sigma^3 \right) \tilde{\eta}\ .
\end{eqnarray}
Note that, when $\delta \psi_r =0$, the remaining condition is the expected Killing spinor equation on $S^2$. As a result it imposes no condition. 

Finally, the dilatino variation can be recast in a  similar fashion to \cite{Itsios:2012dc}: 
\begin{eqnarray}
\label{dil2}
&&  \delta \lambda = e^{X} \left[ \frac{1}{2} \slashed \partial \phi
  - \frac{1}{24}  \slashed{H} \sigma^3 \right] \tilde{\eta} +
 \left[\frac{r^2 + 3 e^{4 A}}{r^2 + e^{4 A}} \Gamma^{r}
 - \frac{2 r e^{2 A}}{r^2+ e^{4 A}} \Gamma^{r \alpha_1\alpha_2 } \sigma^3  \right] \delta \psi_r
\nonumber\\
 &&\phantom{} - e^{X}\left[  \frac{e^{\Phi}}{8} ( 5 m \Gamma^{r} i \sigma^2 + e^{-3 A} \slashed{G}_1 \Gamma^{\alpha_1 \alpha_2} \sigma^1 + 3 \slashed{G}_2 \Gamma^r \sigma^1 +\slashed{G}_3 \Gamma^{\alpha_1 \alpha_2} i \sigma^2 ) \right] \tilde{\eta}\ .
\end{eqnarray}
Neglecting the $\delta \psi_r$ factor, once one redefines the spinors along the lines of (\ref{redefine}), one  realises that the remaining terms are simply the dilatino variation of the original geometry. 

Thus, the essential message of the above analysis is that provided $\delta \psi_r = 0$, we can map these Killing spinor equations back to those of the original geometry. One simply has to guarantee that any conditions arising from $\delta \psi_r$ are consistent with the conditions already imposed on the Killing spinors. 

Therefore, specialising to the geometry of interest to our paper, one can evaluate $\delta \psi_r$ and one encounters a single projection condition 
\begin{equation}
\label{su2proj}
\left[ \cos \theta \Gamma^{\theta r \alpha_1 \alpha_2} \sigma^3 - \sin \theta \Gamma^{\theta r} i \sigma^2 \right] \tilde{\eta} = - \tilde{\eta}. 
\end{equation}
Up to chirality, i.e. through redefinitions (\ref{redefine}), this is simply the projection condition of the original background.  Or to put it another way, once this single condition is imposed, all the Killing spinor equations are satisfied and we conclude that supersymmetry is preserved when one performs a non-Abelian T-duality on the $AdS_6 \times S^4$ solution of massive IIA supergravity. The explicit form of the IIB Killing spinor is 
\begin{equation}
\label{explicitKS}
\eta = e^{X} \tilde{\eta} = e^{X} (\cos \theta)^{-1/12} e^{-\frac{\theta}{2} \Gamma^{\alpha_1 \alpha_2} \sigma^1}  e^{- \frac{\alpha_1}{2} \Gamma^{r \alpha_2} \sigma^3 } e^{- \frac{\alpha_2}{2} \Gamma^{\alpha_2 \alpha_1} } \eta_{AdS_6},
\end{equation}
where we have absorbed all dependence on the $AdS_6$ factor.

\section{Supersymmetric probes}
In this section we complement the DBI analysis presented in the text by exploiting kappa symmetry to confirm that the probes are indeed supersymmetric. We recall that the condition for a probe Dp-brane to be supersymmetric is that it satisfies
\begin{equation}
\Gamma_{\kappa} \eta = \eta, 
\end{equation}
where $\eta$ is the Killing spinor of the background geometry and $\Gamma_{\kappa}$ is a projection matrix,  expressible in the notation of  \cite{Cederwall:1996ri, Bergshoeff:1996tu},  as
\begin{equation}
\label{kappa}
\Gamma_{\kappa} = \frac{\sqrt{|g|}}{\sqrt{|g+ \mathcal{F}}|} \sum_{n=0} \frac{1}{2^n n!} \gamma^{j_1 k_1 \dots j_n k_n} \mathcal{F}_{j_1 k_1} \dots \mathcal{F}_{j_n k_n} J^{(n)}_{(p)}, 
\end{equation}
where $p$ refers to the probe Dp-brane, $\mathcal{F}$ is a combination of Born-Infeld two-form field strength, $F$, and the background NS two-form, $B_2$, $\mathcal{F} = 2 \pi F - B_2$, $g$ denotes the determinant of the induced world-volume metric, $g+\mathcal{F}$ is the latter including $F, B_2$, and finally $\gamma^{i}$ denote induced world-volume gamma matrices. Furthermore, $J^{(n)}_{(p)}$ depends on the Dp-brane probe and $n$ 
\begin{equation}
J^{(n)}_{(p)} =  \biggl\{ \begin{array}{l} (\Gamma_{11})^{n + \frac{p-2}{2}} \Gamma_{(0)}, \\ (-1)^n (\sigma^3)^{n+\frac{p-3}{2}} i \sigma^2 \Gamma_{(0)},  \end{array}
\end{equation}
where upper and lower entries on the RHS distinguish IIA and IIB probes and the matrix $\Gamma_{(0)}$ is given by 
\begin{equation} 
\Gamma_{(0)} = \frac{1}{(p+1)! \sqrt{|g|}} \epsilon^{i_1 \dots i_{(p+1)}} \gamma_{i_1 \dots i_{(p+1)}}. 
\end{equation}

\noindent 
\textbf{D5-brane probes} \\
 For D5-branes wrapping the Minkowski directions $\mathbb{R}_{1,4}$ and the $r$-direction, the $\kappa$-symmetry matrix takes the simple form 
 \begin{eqnarray}
\Gamma^{(r)}_{\kappa} &=& \Gamma_{01234r} \, \sigma^1. 
 \end{eqnarray}
Referring the reader to the explicit form for the background Killing spinor quoted in the text  (\ref{explicitKS}), we note that this projection condition anti-commutes with the $\Gamma^{\alpha_1 \alpha_2} \sigma^3$ term appearing in $e^X$, which can be set to zero provided $\theta = 0$.  If instead of the $r$-direction, the D5-brane wraps the $\theta$-direction, it is not possible to have supersymmetry for finite $r$. \\

\noindent
\textbf{D7-brane probes} \\
Here we consider D7-branes wrapping the Minkowski $\mathbb{R}_{1,4}$, the R symmetry $S^2$ and either $r$ or $\theta$ in the absence of large gauge transformations\footnote{In the presence of large gauge transformations, an additional projector appears that anti-commutes with the $\Gamma^{\alpha_1 \alpha_2} \sigma^3$ term appearing in the $e^{X}$ factor in the Killing spinor. One can reconcile this projector only when $ \theta = 0$.}. Here the $B$-field pulls back to the world-volume of the brane, so we get the following $\kappa$-symmetry projection conditions: 
\begin{eqnarray}
\Gamma^{(r)}_{\kappa} &=& -\Gamma_{01234 r}\, \sigma^1 \,e^{-2X}, \quad \Gamma^{(\theta)}_{\kappa} = - \Gamma_{01234 \theta} \, \sigma^1  \, e^{-2X}, 
\end{eqnarray}
where we have made use of the rotation introduced earlier (\ref{rotate}). These gamma matrices act on the Killing spinors $\eta = e^{X} \tilde{\eta}$ and a pleasing cancellation means that the $e^{X}$ factors drop out so that the $\kappa$-symmetry conditions become respectively 
\begin{equation}
\Gamma_{01234r} \, \sigma^1 \tilde{\eta} =  \tilde{\eta}, \quad \Gamma_{01234 \theta} \, \sigma^1 \tilde{\eta} = \tilde{\eta}. 
\end{equation}
One notices that both of these commute with the projection condition (\ref{su2proj}). However, when the explicit form of the Killing spinor is used (\ref{explicitKS}) one sees that the first projector commutes through the various exponentials, whereas the second projector, corresponding to a D7-brane wrapping $\theta$, requires that $\alpha_1 = 0$. This contradicts the assumption that the $S^2$ is wrapped. As a result, only the D7-brane wrapping the $r$-direction is supersymmetric, a property it possesses for $\theta \in [0, \frac{\pi}{2} ]$.

\end{document}